# Massive Interacting Yang-Mills Multiplets in Nine and Five Dimensions

Hitoshi Nishino[1)] and Subhash Rajpoot[2)]

*Department of Physics & Astronomy*
*California State University*
*1250 Bellflower Boulevard*
*Long Beach, CA 90840*

## Abstract

We present interacting massive $N=1$ vector multiplet (VM) in nine dimensions (9D). Due to the identically-vanishing mass-term $m(\overline\lambda\lambda) \equiv 0$ for (symplectic) pseudo-Majorana gaugino in 9D, we employ unconventional technique to give masses to fermions. In 9D, we consider the gauge group $G$ for the VM $(A_\mu{}^I, \lambda^I, \varphi^I)$ $(I = 1, 2, \cdots, \dim G)$, where $G$ is the Yang-Mills gauge group, and the gaugino $\lambda^I$ is a pseudo-Majorana spinor. We break $G$ by shifting the scalar $\varphi^I$, so that the gaugino $\lambda^I$ as well as its super-partner gauge boson $A_\mu{}^I$ will get the same mass. The scalar $\varphi^I$ plays the role of a Nambu-Goldstone boson absorbed into the longitudinal components of $A_\mu{}^I$, making the latter massive as a super-Proca-Stueckelberg mechanism. We also show that a similar method can be also applied to $N=2$ VMs in 5D.

[1)] E-Mail: hnishino@csulb.edu
[2)] E-Mail: subhash.rajpoot@csulb.edu

# 1. Introduction

One of the most important subjects in supersymmetry is how to give masses to fermions. One method associated with dimensional reduction has been known for many years [1]. It is also well-known that only *pseudo*-Majorana fermions, but *no* Majorana fermions exist in Minkowskian nine-dimensions (9D) with the signature $D = 1+8$ [2][3] *Pseudo*-Majorana spinors, however, allow *no* mass-terms in $D = 1+8$ [2][3]. This forbids the conventional formulation of massive vector multiplets (VMs) in 9D. To be more specific, the naïve mass-term $m(\overline{\psi}\psi)$ for a single pseudo-Majorana fermion $\psi$ in 9D is *identically zero*, because the charge conjugation matrix $C_{\alpha\beta}$ in 9D is *symmetric* [2][3], forcing the naïve mass-term to vanish *identically*: $m(\overline{\psi}\psi) \equiv 0$.

The mass-term problem for a pseudo-Majorana spinor is not just the vanishing lagrangian mass-term $m(\overline{\psi}\psi) \equiv 0$, but it also pops up in the free-field equation. In fact, the massive pseudo-Majorana field equation in 9D is supposed to be $\displaystyle{\not{\partial}}\psi \doteq im\psi$,[3]) because the Clifford algebra in $D=8+1$ [3] requires the imaginary unit in the relative ratio between the two terms in $\displaystyle{\not{\partial}}\psi \doteq im\psi$.[4]) Now the problem is that the resulting Klein-Gordon equation has a *tachyonic* mass:

$$\partial_\mu^2 \psi = \not{\partial}\not{\partial}\psi \doteq \not{\partial}(im\psi) = im\not{\partial}\psi \doteq im(im\psi) = -m^2\psi \quad \Longrightarrow \quad \partial_\mu^2 \psi \doteq -m^2 \psi \qquad (1.1)$$

in our signature $(-,+,+,\cdots,+)$.[5])

The problem in 5D is also similar. In 5D, we have a symplectic spinor $\psi^A$ with the index $A$ for the **2** of $Sp(1)$. Except for the $Sp(1)$ index, the fermionic field equation is $\not{\partial}\psi^A \doteq im\psi^A$ [6]) which is formally the same as the 9D, if the $Sp(1)$ index is suppressed, so that we get again (1.1) with the tachyonic mass,

There have been considerable works related to supersymmetric VMs, such as those in 5D [8][9][10], in dimensions $D \leq 6$ [11], with harmonic-superspaces in 5D [12][13], and likewise in 4D [14]. However, these works never addressed the aforementioned-issue of fermionic mass-term with broken gauge-symmetry in 9D. For example, the papers [8][9][10] focus mainly on the *purely-bosonic* terms. In particular, [8] gives a fermionic propagator in 5D in the Pauli metric $(-,+,+,+,+)^{7)}$ as $1/(\Gamma\cdot p - m)$ *without* the imaginary unit '$i'$' between $\Gamma\cdot p$ and $m$. This is consistent with our (1.1). However, as described above, the trouble is that this leads to the *tachyonic* mass-term. In addition to this, it is *not* clear in [8] that the mass-term in $D=1+4$ needs the non-diagonal $Sp(1)$ metric $\epsilon_{AB}$.

---

[3]) Our space-time signature is $(\eta_{\mu\nu}) = $ diag. $(-,+,+,\cdots,+)$. We also use the symbol $\doteq$ for a field equation distinguished from simply-algebraic ones.

[4]) See [4] and also Appendices B and C for more details.

[5]) We are grateful to P. Townsend for important discussions [5].

[6]) These hermitian properties with or without the imaginary unit are also consistent with $N=1$ and $N=2$ supergravities in 9D [6][7]. See section 4 for details.

The papers [11][12][13] deal only with off-shell or harmonic-superspace formulations, but they never mentioned the case of *massive* VMs, accompanying the gauge-symmery *breaking*, as we perform in this paper. Even though [14] deals with harmonic-superspace in extended $N=2$ supersymmetry within 4D (*not* 5D), giving the mass-term (4.48) in [14], the issue of fermionic mass-term in 9D with *broken* gauge-symmetry has *never* been addressed. Since the fermionic feature in 9D like the *tachyonic* feature is entirely different from 4D, the result in [14] does *not* resolve the problem. The the fermionic structure in 4D [14] is entirely different from 5D, not to mention 9D, and therefore this is irrelevant to our objective. Additionally, since 9D has *neither* off-shell *nor* harmonic-superspace formulation, the results in [11][12][13][14] are *not* of much help. Our main objective is to give the resolution to the *tachyonic* fermion-mass issue in 9D with gauge-symmetry breakings. Even though we will deal with the 5D case, it will be only an additional application of our 9D result.

The trouble with the *tachyonic* mass-term for a single pseudo-Majorana spinor is independent of the absence of a lagrangian mass-term. In other words, this trouble arises not only as an identically-vanishing lagrangian mass-term, but also as a tachyonic mass at the *field-equation* level. On the other hand, according to the general light-cone gauge analyses in diverse dimensions [15], there must be consistent massive VMs in 9D and 5D. From this viewpoint, finding the right formulation of massive VMs must be a technical problem to be solved by setting up the right mass-terms.

In this paper, we overcome the problem of massive VMs in 9D (and apply its technique to 5D). We develop a technique of antisymmetric pairing of gaugino, by shifting the scalar $\varphi^I$ in the VM. The key point is that by the shift $\widetilde{\varphi}^I \equiv \varphi^I + g^{-1}m^I$ by the mass constant $m^I$, there arises the antisymmetric mass-term $f^{IJK}m^I(\overline\lambda^J\lambda^K)$ from the Yukawa-coupling $f^{IJK}(\overline\lambda^I\lambda^J)\varphi^K$. This formalism is interpreted as the Proca-Stueckelberg-like mechanism [16]. As a consequence, the original independent scalar $\varphi^i$ is absorbed into the longitudinal component of $A_\mu{}^I$, making the latter massive. Interestingly, the resulting antisymmetric mass matrix always yields non-tachyonic positive-definite eigenvalues for (mass)$^2$ after the gauge symmetry breaking, as desired.

---

[7)] This is clear in the field equation $(i\rlap/\partial - m)\psi = 0$ in the 18 lines above (2.2) in the published version of [8].

Generally speaking, the supersymmetric formulation of Proca-Stueckelberg mechanism for non-Abelian gauge groups is *not* new. For example, in our recent papers [17], we have presented series of formulations of supersymmetric Proca-Stueckelberg mechanisms in 4D. However, the mechanism we present in this paper has subtle difference in coupling structures, as will be seen in our lagrangians. Thus, our formulation in this paper will provide yet another important example of supersymmetric Proca-Stueckelberg mechanism in 9D or 5D, providing masses to gauge fields.

This paper is organized as follows: In the next section, we start with the lagrangian for $N=1$ massless VM in 9D. In section 3, we introduce the technique to induce masses for the VM. We next analyze the mass spectrum, confirming the Proca-Stueckelberg-like mechanism. In section 4 we apply this formulation to the 5D case, where the only difference is that the pseudo-Majorana gaugino carries an additional $Sp(1)$ index. The concluding remarks are given in section 5. Appendix A is for the general properties of fermions in arbitrary space-time dimensions, while Appendices B is for the applications to 9D.

## 2. N = 1 Superinvariant Action

For $N=1$ supersymmetry in $D=1+8$ with the signature $(+,\cdots,+,-)$, fermions are pseudo-Majorana spinors [2][3]. The hermiticities of fermionic bilinears (BLs) [3][8)] for the pseudo-Majorana fermions $\psi$ and $\chi$ are generalized as $(\overline{\psi}\gamma^{[n]}\chi)^\dagger = -(-1)^n(\overline{\psi}\gamma^{[n]}\chi)$, while the flipping property [3] is $(\overline{\psi}\gamma^{[n]}\chi) = -(-1)^{n(n-1)/2}(\overline{\chi}\gamma^{[n]}\psi)$. Here $\gamma^{[n]}$ ($n = 0, 1, \cdots, 9$) stand for totally antisymmetrized products of $\gamma$-matrices, *e.g.*, $\gamma^{[3]}$ is equivalent to $\gamma^{\mu\nu\rho}$.[9)]

The VM in 9D has the field content $(A_\mu{}^I, \lambda^I, \varphi^I)$, where $I = 1, 2, \cdots, g \equiv \dim G$ are the adjoint index of a non-Abelian gauge group $G$. The total action $I_{9\mathrm{D}} \equiv \int d^9x\, \mathcal{L}_{9\mathrm{D}}$ has the lagrangian[10)]

$$\mathcal{L}_{9\mathrm{D}} = -\tfrac{1}{4}(F_{\mu\nu}{}^I)^2 + \tfrac{1}{2}(\overline{\lambda}^I \slashed{D} \lambda^I) - \tfrac{1}{2}(D_\mu \varphi^I)^2 - \tfrac{i}{2}gf^{IJK}(\overline{\lambda}^I \lambda^J)\varphi^K \quad , \tag{2.1}$$

where $g$ is the gauge-coupling. The field strength and the covariant derivatives are

$$F_{\mu\nu}{}^I \equiv +2\partial_{[\mu} A_{\nu]}{}^I + g f^{IJK} A_\mu{}^J A_\nu{}^K \quad , \tag{2.2a}$$

---

[8)] We can also refer the readers to [7] with the caveat about the space-time signature difference.

[9)] For more details, see Appendix A & B.

[10)] The validity of the presence or absence of the imaginary unit in the metric $(-,-,\cdots,-,+)$ is easily re-confirmed with 9D supergravity [5]. For the $(-,+,+,\cdots,+)$-metric, see [7].

$$\mathcal{D}_\mu \lambda^I \equiv +\partial_\mu \lambda^I + g f^{IJK} A_\mu{}^J \lambda^K ~, \qquad (2.2\text{b})$$

$$\mathcal{D}_\mu \varphi^I \equiv +\partial_\mu \varphi^I + g f^{IJK} A_\mu{}^J \varphi^K ~. \qquad (2.2\text{c})$$

The total action $I_{9\text{D}}$ is invariant under $N=1$ supersymmetry

$$\begin{aligned}
\delta_Q A_\mu{}^I &= -(\overline{\epsilon}\gamma_\mu \lambda^I) ~, \\
\delta_Q \lambda^I &= -\tfrac{1}{2}(\gamma^{\mu\nu}\epsilon) F_{\mu\nu}{}^I - i(\gamma^\mu \epsilon)\mathcal{D}_\mu \varphi^I ~, \\
\delta_Q \varphi^I &= -i(\overline{\epsilon}\lambda^I) ~,
\end{aligned} \qquad (2.3)$$

The commutator algebra for two supersymmetry transformations is

$$[\delta_Q(\epsilon_1), \delta_Q(\epsilon_1)] = \delta_P(\xi) + \delta_T(\alpha) ~,$$
$$\xi^\mu \equiv +2(\epsilon_1 \gamma^\mu \epsilon_2) ~, \qquad \alpha^I \equiv -\xi^\mu A_\mu{}^I + 2i(\overline{\epsilon}_2 \epsilon_1)\varphi^I ~, \qquad (2.4)$$

where $\delta_T$ is the $G$-group gauge transformation with the parameter $\alpha^I$. Note that there is *no* central charge involved at this stage. However, this situation changes, when we consider the massive case in section 4. Since our formulation is *on-shell* formulation, the commutator algebra closes by the use of $\lambda$ and $\chi$-field equations[11]

$$\frac{\delta \mathcal{L}_{9\text{D}}}{\delta \overline{\lambda}^I} = +(\not{\mathcal{D}} \lambda^I) - i g f^{IJK} \lambda^J \varphi^K \doteq 0 ~, \qquad (2.5)$$

### 3. Mass Generation

The technique to create non-tachyonic masses for the fields in the VM $(A_\mu{}^I, \lambda^I, \varphi^I)$ is as follows. We shift the scalar field $\varphi$ to $\widetilde{\varphi}$ as

$$\varphi^I \equiv \widetilde{\varphi}^I - g^{-1} m^I ~, \qquad (3.1)$$

where $m^I$ are constants with dimension of mass. As is well known, this shift induces the mass-term in the $A_\mu{}^I$-kinetic term. In fact, the original lagrangian (2.1) becomes now

$$\begin{aligned}
\mathcal{L}_{9\text{D}} =~& -\tfrac{1}{4}(F_{\mu\nu}{}^I)^2 + \tfrac{1}{2}(\overline{\lambda}^I \not{\mathcal{D}} \lambda^I) \\
& -\tfrac{1}{2}\left(\partial_\mu \widetilde{\varphi}^I + g f^{IJK} A_\mu{}^J \widetilde{\varphi}^K + f^{IJK} m^J A_\mu{}^K \right)^2 \\
& -\tfrac{i}{2} g f^{IJK} (\overline{\lambda}^I \lambda^J) \widetilde{\varphi}^K + \tfrac{i}{2} f^{IJK} m^I (\overline{\lambda}^J \lambda^K)
\end{aligned} \qquad (3.2)$$

---

[11] We use the symbol $\doteq$ meant for a field equation, distinguished from a merely algebraic equality.

First, the last term in (3.2) can be regarded as the gaugino mass-term. If we regard $\widetilde{\varphi}^I$ as a new independent field, the $\lambda$-field equation is now

$$\displaystyle{\not}\!{D} \lambda^I \doteq -i\mathcal{M}^{IJ}\lambda^K + \mathcal{O}(\phi^2) \ , \qquad (\mathcal{M}^{IJ} \equiv f^{IJK}m^K = -\mathcal{M}^{JI}) \ . \tag{3.3}$$

Here $\mathcal{O}(\phi^2)$ implies all quadratic terms for interactions, and the matrix $\mathcal{M}^{IJ}$ is a $d$ by $d$ *antisymmetric* matrix for $d \equiv \dim G$. In order to determine whether the gaugino is tachyonic, we analyze the Klein-Gordon equation by

$$\begin{aligned} \partial_\mu^2 \varphi^I = \displaystyle{\not}\!{\partial}(\displaystyle{\not}\!{\partial}\lambda^I) &\doteq \displaystyle{\not}\!{\partial}(-i\mathcal{M}^{IJ}\lambda^J) + \mathcal{O}(\phi^2) = -i\mathcal{M}^{IJ}\displaystyle{\not}\!{\partial}\lambda^J + \mathcal{O}(\phi^2) \\ &= -i\mathcal{M}^{IJ}(-i\mathcal{M}^{JK}\lambda^K) + \mathcal{O}(\phi^2) = -(\mathcal{M}^2)^{IJ}\lambda^J + \mathcal{O}(\phi^2) \ . \end{aligned} \tag{3.4}$$

The question now is what are the eigenvalues of the mass matrix $-(\mathcal{M}^2)^{IJ}$. The answer depends on whether $d \equiv \dim G$ is an even or odd integer. We can confirm the facts that

**(i)** If $d = (\text{even}) \equiv 2k$ $(k = 1, 2, 3, \cdots)$, all eigenvalues of $(-\mathcal{M}^2)$ are *positive-definite* real numbers.

**(ii)** If $d = (\text{odd}) \equiv 2k-1$ $(k = 1, 2, 3, \cdots)$, at least one eigenvalue of $(-\mathcal{M}^2)$ is zero, while all other eigenvalues are *positive-definite* real numbers.

These statements are confirmed as follows: For the case (i), we know that arbitrary real anti-symmetric real matrix $\mathcal{M}$ is diagonalized to $\mathcal{M}_{\text{d}}$ by a unitary matrix $U$ as [18]

$$\mathcal{M}_{\text{d}} = U^{-1}\mathcal{M}U = \text{diag.}(+i\nu_1, -i\nu_1, +i\nu_2, -i\nu_2, \cdots, +i\nu_k, -i\nu_k) \ , \tag{3.5}$$

where $\nu_i$ $(i = 1, 2, \cdots, k)$ are all real numbers. This is because the original matrix $\mathcal{M}$ is real, so that when it is diagonalized all of its eigenvalues are pure imaginary, paired up as complex conjugates: $\pm i\nu_1, \pm i\nu_2, \cdots, \pm i\nu_k,$ as in (3.5). This property has been also related to the so-called 'Pfaffian' [19]

$$\det \mathcal{M} = \nu_1^2 \cdot \nu_2^2 \cdots \cdot \nu_k^2 = [\,\text{pf}(\mathcal{M})\,]^2 \geq 0 \ . \tag{3.6}$$

Eq. (3.5) implies that the matrix $-\mathcal{M}^2$ has *positive-definite*[12] real-number eigenvalues, and is diagonalized as

$$-(\mathcal{M}^2)_{\text{d}} = -(\mathcal{M}_{\text{d}})^2 = \text{diag.}\left(+\nu_1^2, +\nu_1^2, +\nu_2^2, +\nu_2^2, \cdots, +\nu_k^2, +\nu_k^2\right) \quad (Q.E.D.) \tag{3.7}$$

---

[12] The phrase '*positive-definite*' includes the case of *accidental zeros* among $\nu_i$ $(i = 1, 2, \cdots, k)$.

For the case (ii), we follow the Jacobi's theorem [20] that a $(2k-1)$ by $(2k-1)$ anti-symmetric matrix has a vanishing determinant:

$$\det \mathcal{M} = \det(\mathcal{M}^T) = \det(-\mathcal{M}) = (-1)^{2k-1} \det \mathcal{M} = -\det \mathcal{M} = -(\text{LHS})$$
$$\implies \quad \det \mathcal{M} = 0 \ . \tag{3.8}$$

It then follows that at least one eigenvalue of the matrix $\mathcal{M}$ is zero. As for the remaining eigenvalues, it is similar to the case (i): $d = 2k$, namely, all these eigenvalues are pure imaginary, and paired up as complex conjugates. Therefore,

$$\mathcal{M}_\text{d} = U^{-1} \mathcal{M} U = \text{diag.}(+i\nu_1, -i\nu_1, +i\nu_1, -i\nu_1, \cdots, +i\nu_\ell, -i\nu_\ell, \overbrace{0, \cdots, 0}^{2k-2\ell-1})$$
$$\implies \quad -(\mathcal{M}^2)_\text{d} = -(\mathcal{M}_\text{d})^2$$
$$= \text{diag.}\left(+\nu_1^2, +\nu_1^2, +\nu_2^2, +\nu_2^2, \cdots, +\nu_\ell^2, +\nu_\ell^2, \overbrace{0, \cdots, 0}^{2k-2\ell-1}\right) . \tag{3.9}$$

where $1 \leq \ell \leq k$. In other words, at least one eigenvalue of the matrix $(-\mathcal{M}^2)$ is zero, while all other eigenvalues are *positive-definite* real numbers (*Q.E.D.*)

Applying these results to the mass operator in (3.3), we can conclude that all the eigenvalues of the mass operator $(-\mathcal{M}^2)^{IJ}$ are *positive-definite*, so that there arises *no tachyonic mass* for the gaugino. The basic mechanism is easily understood as follows. The pairing of $\pm i\nu_i$ eigenvalues is equivalent to a Jordan block by the 2 by 2 anti-symmetric matrix $m\epsilon^{ij}$, $\epsilon^{12} = -\epsilon^{21} = +1$, $\epsilon^{11} = \epsilon^{22} = 0$, so the basic block of the BL-term of the $\lambda$-equation (3.3) has the structure

$$\partial\!\!\!/\chi^i \doteq -m\epsilon^{ij}\chi^j \tag{3.10}$$

where we have replaced $\lambda^I$ by $\chi^i$ ($i$ = 1. 2) for each 2 by 2 Jordan block, omitting also the adjoint index. This leads to the Klein-Gordon equation with *non-tachyonic mass*:

$$\partial_\mu^2 \chi^i = \partial\!\!\!/(\partial\!\!\!/\chi^i) \doteq \partial\!\!\!/(-im\epsilon^{ij}\chi^j) = -im\epsilon^{ij}\partial\!\!\!/\chi^j = -im\epsilon^{ij}(-im\epsilon^{jk}\chi^k) = +m^2\chi^i \ . \tag{3.11}$$

In other words, the doubling within each 2 by 2 block resolves the tachyonic-mass problem in (1.1).

Since supersymmetry is *unbroken*, we can expect similar non-tachyonic masses for the gauge boson $A_\mu{}^I$. As a matter of fact, this is manifestly seen as follows. The BL-terms of bosons in the lagrangian (3.2) are

$$\mathcal{L}_\text{Bos, BL} = -\tfrac{1}{4}(F_{\mu\nu}{}^I)^2 - \tfrac{1}{2}\left(\partial_\mu \varphi^I - \mathcal{M}^{IJ} A_\mu{}^J\right)^2 \ . \tag{3.12}$$

We ignored trilinear or higher-order interaction terms. In order to study the mass-terms in (3.12), we also need to eliminate the BL-order mixture between $A_\mu{}^I$ and $\varphi^I$. To this end, we limit ourselves to the special case of $d = $ (even) $\equiv 2k$, and assume that *all eigenvalues of the matrix $\mathcal{M}^{IJ}$ are non-zero*. It the follows that *all eigenvalues of the matrix $(-\mathcal{M}^2)^{IJ}$ are positive*,[13)] and therefore, its inverse matrix $\mathcal{M}^{-1}$ exists:

$$(\mathcal{M}^{-1})^{IJ} \mathcal{M}^{JK} = \delta^{IK} \ . \tag{3.13}$$

Using $\mathcal{M}^{-1}$, it is straightforward to eliminate the BL-order mixture between $A_\mu{}^I$ and $\varphi^I$ by the field redefinition

$$\widetilde{A}_\mu{}^I \equiv A_\mu{}^I - (\mathcal{M}^{-1})^{IJ} \partial_\mu \widetilde{\varphi}^J \tag{3.14}$$

leading to

$$\mathcal{L}_{\text{Bos, } \phi^2} = -\tfrac{1}{4}(\widetilde{F}_{\mu\nu}{}^I)^2 + \tfrac{1}{2}(\mathcal{M}^2)^{IJ} \widetilde{A}_\mu{}^I \widetilde{A}^{\mu J} + \mathcal{O}(\phi^3) \ . \tag{3.15}$$

Here $\widetilde{F}_{\mu\nu}{}^I$ is the same as (2.2a), except that $A_\mu{}^I$ is now replaced by $\widetilde{A}_\mu{}^I$. The $\widetilde{F}_{\mu\nu}{}^I$ is *not* exactly the same as $F_{\mu\nu}{}^I$, but the difference arises at higher-order terms containing $\widetilde{\varphi}$, but they do not interest us at this stage. As we have seen, (3.15) implies *non-tachyonic* mass for $A_\mu{}^I$, because of the *positive-definiteness* of all eigenvalues of $(-\mathcal{M}^2)^{IJ}$. This result is also consistent with the mass spectrum for the gaugino $\lambda^I$ in (3.7), as desired for a supersymmetric partner. In other words, $N = 1$ supersymmetry is maintained in our mechanism.

We emphasize that our mechanism of providing a mass-matrix to the gauge field is interpreted as Proca-Stueckelberg (compensator) mechanism, consistent also with $N = 1$ supersymmetry in 9D. In this sense, our formulation is providing yet another example of supersymmetric compensator mechanism for non-Abelian gauge group. Due to the different fermionic structure, this new mechanism is is different from supersymmetric compensator mechanism in 4D [17].

We mention the effect of the mass-term in (3.1) on the commutator algebra (2.3). As is easily seen, the mass-term generates the new term as

$$\alpha^I \equiv -\xi^\mu A_\mu{}^I + 2i(\overline{\epsilon}_2 \epsilon_1)\varphi^I + 2ig^{-1} m^I (\overline{\epsilon}_1 \epsilon_2) \ . \tag{3.16}$$

The last term is interpreted as nothing but the central charge in 9D, predicted from the general algebraic argument in [15].

---

[13)] Our assumption excludes even *accidental* zero eigenvalue.

Note that our mass-generation mechanism itself does *not* break $N=1$ supersymmetry. Despite the *unbroken* $N=1$ supersymmetry, the original gauge symmetry for the group $G$ has been broken, due to the compensator mechanism, played by the compensator $\varphi^I$. This also explains why the mass matrix $\mathcal{M}^{IJ}$ depends on the adjoint indices $IJ$, which obviously breaks the original gauge symmetry.

## 4. Parallel Structures for 5D Case

The result and method for our 9D case can be applied to 5D with $N=2$ supersymmetry.

Since the most of the notation for $N=2$ supersymmetry in 5D has been well-known in [21], we skip their details. Instead of *pseudo*-Majorana spinors, we have *symplectic pseudo*-Majorana spinors [3][2]. The relevant multiplets are the VM $(A_\mu{}^I, \lambda^{AI}, \varphi^I)$ in 5D, where $\lambda^{AI}$ is a $Sp(1)$ symplectic pseudo-Majorana spinors with additional index $A=1,2$ for the **2** of $Sp(1)$ [3][2][21][22]. Accordingly, their BLs need additional contractions with the $Sp(1)$ metric $\epsilon_{AB}$ which are sometimes omitted, such as $(\overline{\chi}\slashed{D}\lambda) \equiv (\overline{\chi}^A\slashed{D}\lambda_A) \equiv (\overline{\chi}^A\slashed{D}\lambda^B)\epsilon_{BA}$, etc. The hermiticities of Majorana BLs are $(\overline{\chi}\gamma^{[n]}\lambda)^\dagger \equiv (\overline{\chi}^A\gamma^{[n]}\lambda_A)^\dagger = -(-1)^n(\overline{\chi}\gamma^{[n]}\lambda)$ [21][3][22], while the flipping property is $(\overline{\chi}\gamma^{[n]}\lambda) = -(-1)^{n(n-1)}(\overline{\lambda}\gamma^{[n]}\chi)$ [3][2].

The total action is $I_{5D} \equiv \int d^5x \, \mathcal{L}_{5D}$ has a structure similar to the 9D case in (3.1):

$$\mathcal{L}_{5D} = -\tfrac{1}{4}(F_{\mu\nu}{}^I)^2 + \tfrac{1}{2}(\overline{\lambda}^I\slashed{D}\lambda^I) - \tfrac{1}{2}(D_\mu\varphi^I)^2 - \tfrac{i}{2}gf^{IJK}(\overline{\lambda}^I\lambda^J)\varphi^K \quad , \tag{4.1}$$

where the field strengths are covariant derivatives are defined in the same way as (2.2). The action $I_{5D}$ is invariant under $N=2$ supersymmetry

$$\delta_Q A_\mu{}^I = +(\overline{\epsilon}\gamma_\mu\lambda^I) \quad , \quad \delta_Q\varphi^I = +i(\overline{\epsilon}\lambda^I) \quad ,$$
$$\delta_Q\lambda^I = -\tfrac{1}{2}(\gamma^{\mu\nu}\epsilon)F_{\mu\nu}{}^I + i(\gamma^\mu\epsilon)D_\mu\varphi^I \quad . \tag{4.2}$$

The $\lambda^I$-field equation is simply

$$\slashed{D}\lambda^I - igf^{IJK}\lambda^J\varphi^K \doteq 0 \quad . \tag{4.3}$$

After the same shift as (3.1), we get

$$\slashed{D}\lambda^I - i\mathcal{M}^{IJ}\lambda^J - igf^{IJK}\lambda^J\widetilde{\varphi}^K \doteq 0 \quad . \tag{4.4}$$

The second term is the mass-term with the same definition (3.3) for $\mathcal{M}^{IJ}$, while the last term is an interaction term at $\mathcal{O}(\phi^2)$. As in the previous 9D case, this leads to the *non-tachyonic*

Klein-Gordon mass:

$$\partial_\mu^2 \lambda^I = \not{\partial}(\not{\partial}\lambda^I) \doteq \not{\partial}(i\mathcal{M}^{IJ}\lambda^I) + \mathcal{O}(\phi^2) = i\mathcal{M}^{IJ}(\not{\partial}\lambda^I) + \mathcal{O}(\phi^2)$$
$$= i\mathcal{M}^{IJ}(i\mathcal{M}^{JK}\lambda^K) + \mathcal{O}(\phi^2) = -(\mathcal{M}^2)^{IJ}\lambda^J + \mathcal{O}(\phi^2) \ . \tag{4.5}$$

Again the mass matrix $-(\mathcal{M}^2)^{IJ}$ with *positive-definite* eigenvalues arises, guaranteeing the *absence of tachyonic mass*. Note that this mechanism is essentially the same as in 9D, despite the presence or absence of imaginary unit '$i$' caused by the notational difference from 9D.

As for the BL-order mixture between $\widetilde\varphi^I$ and $A_\mu{}^I$, its mechanisms is also parallel to the 9D case. Therefore we sill skip their details here.

## 5. Concluding Remarks

In this paper, we have presented the formulation of massive VMs with non-trivial interactions in 9D and 5D. We have solved the problem of vanishing or tachyonic mass-terms for *pseudo*-Majorana spinors with non-trivial interactions. We have introduced the technique of inducing *non-tachyonic* masses for the VMs, resolving the usual mass-term problem for VM with *pseudo*-Majorana spinors for a general gauge group $G$.

The properties of mass matrix $\mathcal{M}^{IJ}$ are associated with the non-diagonal mass-term inherent to *pseudo*-Majorana spinors in 9D. The antisymmetric property of the matrix $\mathcal{M}^{IJ}$ is closely related to the property of the Pfaffian, and is in turn related to the *positive-definite eigenvalues* of the mass matrix $(-\mathcal{M}^2)^{IJ}$, which imply the *non-tachyonic* masses for the VM.

In the case of $g \equiv \dim G = (\text{odd}) \equiv 2k - 1$, at least one eigenvalue of the mass matrix $(-\mathcal{M}^2)^{IJ}$ is zero. This further means that some component(s) among $A_\mu{}^I$ stay massless, and therefore, *no* symmetry-breaking occurs for certain generators, with unbroken $U(1) \cong SO(2)$ symmetry. Note also that $N=1$ (or $N=2$) supersymmetry is maintained *unbroken* in 9D (or 5D).

We have seen that our original problems with mass-terms for the *pseudo*-Majorana spinors in 9D or 5D have been solved in terms of anti-symmetric mass matrix $\mathcal{M}$. This property seems to be peculiar to 9D or 5D, because we did *not* encounter similar properties in other dimensions, such as 4D, where *diagonal* mass-terms are allowed like $m(\overline\lambda^I \lambda^I) \neq 0$.

Since our formulation is based on the antisymmetry of the structure constant $f^{IJK}$, our conclusion is valid for any classical compact groups, such as $A_n \equiv SU(n+1)$, $B_n \equiv$ $SO(2n+1)$, $C_n \equiv Sp(2n)$, $D_n \equiv SO(2n)$, as well as exceptional compact groups $G_2$, $F_4$, $E_6$, $E_7$ and $E_8$. Depending on whether $d =$ (even) or $d =$ (odd), the breaking patterns and mass-spectrum are determined.

As has been also mentioned, our mechanisms for massive gauge fields provide the additional examples of supersymmetric Proca-Stueckelberg (compensator) formulations. To be more specific, the vector-multiplet in 9D is $(A_\mu{}^I, \lambda^I, \varphi^I)$, where the scalar $\varphi^I$ plays the role of a compensator, absorbed into $A_\mu{}^I$ making the latter massive. One important aspect is that while the gauge symmetry for the non-Abelian group $G$ is broken, the original $N = 1$ supersymmetry is *not* broken, showing the consistency of supersymmetric compensator mechanism.

These mechanisms work in odd dimensions such as 5D or 9D, where fermionic structures are different from 4D. From this viewpoint, our results in this paper can play leading roles for exploiting compensator-field formulations for massive gauge fields in higher dimensions in the future.

As the last words, we stress one additional important point in our results. To our knowledge, there has been *no* paper that dealt with gauge-breakings for VMs in higher dimensions, such as $D \geq 9$. In this paper, we have given the non-trivial gauge-breaking mechanism that has *not* been known before in $D \geq 9$. Even though our mechanism is based on the Proca-Stueckelberg-like mechanism [16], it is closely related to the subtlety of the gaugino mass-terms in 9D.

Appendices A and B give useful relationships about fermions, in particular, such as (A.1) through (A.3) with Table A-1 with minor typographical errors in [3] now corrected. These relationships will be of considerable importance for future research associated with fermions in higher dimensions in addition to 9D and 5D, that we have given explicitly in these appendices. We believe that our result in this paper paves the way for further studies of supersymmetric models in higher odd dimensions with interacting VMs in non-adjoint representations.

We are grateful to E. Sezgin who re-confirmed the typographical errors in [3]. We are also indebted to W. Siegel and P. Townsend for valuable discussions. We also acknowledge the referee of this paper for pointing out important references [8] ~ [14] we should *not* overlook.

## Appendix A: Fermions in Diverse Space-Time Dimensions $^\forall D$

In this appendix, we clarify the general properties of fermions in $^\forall D$ space-time dimensions.[14] To this end, we follow the general analysis by Salam-Sezgin [3] in general space-time dimensions $^\forall D$. Consider the Clifford algebra $\{\gamma_\mu, \gamma_\nu\} = +2\eta_{\mu\nu}$ with the metric $(\eta_{\mu\nu}) = \text{diag.}(\overbrace{-,-,-,\cdots,-}^{t}, \overbrace{+,+,\cdots,+}^{s},)$, where $t$ (or $s$) is the number of time (or spatial) dimensions. The properties of gamma-matrices in space-time dimensions $^\forall D = t + s$ with the coordinate index $\mu = 0, 1, \cdots; t-1, t, \cdots, D-1$ are such as

$$(\gamma_0)^\dagger = -\gamma_0 \ , \quad (\gamma_1)^\dagger = -\gamma_1 \ , \quad \cdots \ , \quad (\gamma_{t-1})^\dagger = -\gamma_{t-1} \ , \tag{A.1a}$$

$$(\gamma_t)^\dagger = +\gamma_t \ , \quad (\gamma_{t+1})^\dagger = +\gamma_{t+1} \ , \quad \cdots \ , \quad (\gamma_{D-1})^\dagger = +\gamma_{D-1} \ , \quad \overline{\psi} \equiv \psi^\dagger A \ , \tag{A.1b}$$

$$\gamma_\mu^\dagger = (-1)^t A \gamma_\mu A^{-1} \ , \quad A \equiv \gamma_0 \gamma_1 \cdots \gamma_{t-1} \ , \quad A^\dagger \equiv (-1)^{t(t+1)/2} A \ , \quad B^\dagger B = I \ , \tag{A.1c}$$

$$\gamma_\mu^* = \eta \, B \gamma_\mu B^{-1} \ , \quad C \equiv BA \ , \quad B = CA^{-1} \ , \quad A = B^{-1}C \ , \quad \epsilon^2 = 1 \ , \quad \eta^2 = 1 \ , \tag{A.1d}$$

$$\gamma_\mu^T = (-1)^t \eta \, C \gamma_\mu C^{-1} \ , \quad C^\dagger C = +I \ , \quad C^T = \epsilon \, \eta^t (-1)^{t(t+1)/2} C \ , \quad B^T = \epsilon B \ . \tag{A.1e}$$

The signatures $\epsilon$ and $\eta$ are determined by $s - t$ as in table A-1 [3]:[15]

| $s - t$ | $\epsilon$ and $\eta$ | Kind of Fermions | Condition |
|---|---|---|---|
| 1, 2, 8 (mod 8) | $\epsilon = +1, \eta = +1$ | Majorana | $\psi^* = B\psi$ |
| 6, 7, 8 (mod 8) | $\epsilon = +1, \eta = -1$ | Pseudo-Majorana | $\psi^* = B\psi$ |
| 4, 5, 6 (mod 8) | $\epsilon = -1, \eta = +1$ | Symplectic-Majorana | $\psi^{*A} = (\psi_A)^* = \epsilon^{AB} B \psi_B$ |
| 2, 3, 4 (mod 8) | $\epsilon = -1, \eta = -1$ | Pseudo-Symplectic-Majorana | $\psi^{*A} = (\psi_A)^* = \epsilon^{AB} B \psi_B$ |

Table A-1: Fermions in Diverse Dimensions

Other important properties are such as

$$(\overline{\psi} \gamma^{[n]} \chi) = -\epsilon \eta^{t+n} (-1)^{(t-n)(t-n+1)/2} (\overline{\chi} \gamma^{[n]} \psi) \ , \tag{A.2a}$$

$$(\overline{\psi} \gamma^{[n]} \chi)^\dagger = +\epsilon \eta^{t+n} (\overline{\psi} \gamma^{[n]} \chi) \ , \tag{A.2b}$$

for (pseudo-)Majorana fermions, and

$$(\overline{\psi}^A \gamma^{[n]} \chi_B) = -\epsilon \eta^{t+n} (-1)^{(t-n)(t-n+1)/2} (\overline{\chi}_B \gamma^{[n]} \psi^A) \ , \tag{A.3a}$$

$$(\overline{\psi}^A \gamma^{[n]} \chi_B)^\dagger = +\epsilon \eta^{t+n} (\overline{\psi}_A \gamma^{[n]} \chi^B) \ , \tag{A.3b}$$

---

[14] This appendix corrects crucial typographical errors in [3].

[15] The $\epsilon^{AB}$ in the table is the $Sp(1)$ metric.

for (pseudo-)symplectic Majorana fermions with $Sp(1)$ indices $A, B = 1, 2$. Here the symbol $[n]$ stands for the antisymmetric indices $\mu_1 \cdots \mu_n$ to save space.

**Appendix B: Fermions in $D = 1 + 8$**

As is seen in Table A-1, the case of 9D ($D = 1 + 8$) gives $s - t = 8 - 1 = 7$, uniquely fixing the fermions in $D = 1 + 8$ as *pseudo-Majorana* spinors, with $\epsilon = +1$, $\eta = -1$. According to (A.1e), the charge-conjugation matrix is *symmetric*: $C^T = +C$. Eqs. (A.2a) and (A.2b) for our 9D case are

$$(\overline{\psi}\gamma^{[n]}\chi) = -(-1)^{n(n-1)/2}(\overline{\chi}\gamma^{[n]}\psi) \ , \tag{B.1a}$$

$$(\overline{\psi}\gamma^{[n]}\chi)^\dagger = -(-1)^n(\overline{\psi}\gamma^{[n]}\chi) \ . \tag{B.1b}$$

In 9D, the case of $n = 0$ in (B.1a) leads to $(\overline{\psi}\chi) = -(\overline{\chi}\psi)$, implying the vanishing of the conventional mass-term: $m(\overline{\chi}\chi) \equiv 0$. On the other hand, (B.1b) implies $(\overline{\psi}\chi)^\dagger = -(\overline{\psi}\chi)$, $(\overline{\psi}\gamma^\mu\chi)^\dagger = +(\overline{\psi}\gamma^\mu\chi)$, meaning that the terms $im(\overline{\psi}\chi)$ and $(\overline{\psi}\partial\!\!\!/\chi)$ are hermitian lagrangian terms. This implies that the expected right free massive $\chi$-field equation should be $\partial\!\!\!/\chi \doteq im\chi$, as mentioned for (1.1).

**Appendix C: Fermions in $D = 1 + 4$**

Our notation for fermions in $D = 1 + 4$ coincides with that in [21]. Nevertheless, we give the brief summary of our conventions, in connection with Table A-1.

As is seen in Table A-1, the case of 5D ($s = 4$, $t = 1$) with $s - t = 4 - 1 = 3$ uniquely fixing the fermions in $D = 1 + 4$ as *pseudo-Symplectic Majorana* spinors, with $\epsilon = \eta = -1$. According to (A.1e), the charge-conjugation matrix is *symmetric*: $C^T = +C$. Eqs. (A.3a) and (A.3b) for our 5D case are

$$(\overline{\psi}^A\gamma^{[n]}\chi_B) = +(-1)^{n(n-1)/2}(\overline{\chi}_B\gamma^{[n]}\psi^A) \ , \tag{C.1a}$$

$$(\overline{\psi}^A\gamma^{[n]}\chi_B)^\dagger = +(-1)^n(\overline{\psi}_A\gamma^{[n]}\chi^B) \ . \tag{C.1b}$$

In 5D, the case of $n = 0$ in (C.1a) leads to $(\overline{\psi}\chi) = -(\overline{\chi}\psi)$, implying the vanishing of the conventional mass-term: $m(\overline{\chi}\chi) \equiv 0$. On the other hand, (C.1b) implies $(\overline{\psi}^A\chi_A)^\dagger = -(\overline{\psi}^A\chi_A)$, $(\overline{\psi}^A\gamma^\mu\chi_A)^\dagger = +(\overline{\psi}^A\gamma^\mu\chi_A)$, meaning that the terms $im(\overline{\psi}^A\chi_A)$ and $(\overline{\psi}^A\partial\!\!\!/\chi_A)$ are hermitian lagrangian terms. These also agree with [21]. It also follows that the expected right free massive $\chi$-field equation should be $\partial\!\!\!/\chi \doteq im\chi$, as mentioned in (1.1).

# References

[1] J. Scherk and J.H. Schwarz, Nucl. Phys. **B153** (1979) 61.

[2] T. Kugo and P.K. Townsend, Nucl. Phys. **B221** (1983) 357.

[3] *'Supergravities in Diverse Dimensions'*, *eds.* A. Salam and E. Sezgin (North Holland/World-Scientific 1989), Vol. **1**, page 5.

[4] *See, e.g.,* M. Günaydin. G. Sierra and P.K. Townsend, Nucl. Phys. **B242** (1984) 244; M. Günaydin and M. Zagermann, Phys. Rev. **D62** (2000) 044028, `hep-th/0002228`.

[5] P. Townsend, *private communication*.

[6] H. Nishino and S. Rajpoot, Phys. Lett. **546B** (2002) 261.

[7] S.J. Gates, Jr., H. Nishino and E. Sezgin, Class. & Quant. Gr. **3** (1986) 21

[8] E. Witten, Nucl. Phys. **B471** (1996) 195, `hep-th/9603150`.

[9] N. Seiberg, Phys. Lett. **388B** (1996) 753, `hep-th/9608111`.

[10] K.A. Intriligator, D.R. Morrison and N. Seiberg, Nucl. Phys. **B497** (1997) 56, `hep-th/9702198`.

[11] B. Zupnik, Nucl. Phys. **B554** (1999) 365, Erratum-*ibid.* **B644** (2002) 405, `hep-th/9902038`.

[12] S.M. Kuzenko and W.D. Linch, JHEP **0602** (2006) 038, `hep-th/0507176`.

[13] I.L. Buchbinder and N.G. Pletnev, JHEP **1511** (2015) 130, arXiv:1510.02563[hep-th], *and references therein*.

[14] A. Galperin, E. Ivanov, S. Kalitzin, V. Ogievetsky and E. Sokatchev, Class. & Quant. Gr. **1** (1984) 469.

[15] J. Strathdee, *'Extended Poincare Supersymmetry'*, Int. Jour. Mod. Phys. **A2** (1987) 273.

[16] A. Proca, J. Phys. Radium **7** (1936) 347; E.C.G. Stueckelberg, Helv. Phys. Acta **11** (1938) 225; *See, e.g.*, D. Feldman, Z. Liu and P. Nath, Phys. Rev. Lett. **97** (1986) 021801. *For reviews, see, e.g.*, H. Ruegg and M. Ruiz-Altaba, Int. Jour. Mod. Phys. **A19** (2004) 3265.

[17] H. Nishino and S. Rajpoot, Phys. Rev. **D83** (2011) 085008; Nucl. Phys. **B872** (2013) 213; *ibid.* **B887** (2014) 265.

[18] *See, e.g.,* 'Encyclopedic Dictionary of Mathematics', edited by Kiyosi Ito, The Mathematical Society of Japan (MIT Press, Cambridge, MA, 1987).

[19] A. Cayley, *"On the Theory of Permutants"*. Cambridge and Dublin Mathematical Journal VII: 40, Reprinted in Collected mathematical papers, volume 2.

[20] H. Eves, *'Elementary Matrix Theory'*, Dover Publications, ISBN 978-0-486-63946-8.

[21] M. Günaydin, G. Sierra and P.K. Townsend, Phys. Lett. **133B** (1983) 72; *ibid.* **144B** (1984) 41; Phys. Rev. Lett. **53** (1984) 332; Nucl. Phys. **B242** (1984) 244; *ibid.* **B253** (1985) 573.

[22] H. Nishino and S. Rajpoot, Phys. Lett. **502B** (2001) 246; Nucl. Phys. **B612** (2001) 98.